\documentclass[a4paper,10pt]{article}

\usepackage[hmargin=35mm,vmargin=30mm]{geometry}

\usepackage{amsmath}              % For mathematical symbols
\usepackage{amssymb}              % For mathematical symbols
\usepackage{amsthm}               % For custom theorem styles
\usepackage{enumerate}            % To support \begin{enumerate}[(i)]
\usepackage{graphicx}             % For embedding figures
\usepackage[loose]{subfigure}     % For embedding subfigures
\usepackage[noend]{algpseudocode} % For formatting pseudocode within algorithms
\usepackage{pictexwd,dcpic}       % For two-dimensional lattices

\newtheorem{theorem}{Theorem}[section]
\newtheorem{lemma}[theorem]{Lemma}
\newtheorem{corollary}[theorem]{Corollary}
\newtheorem{problem}[theorem]{Problem}
\newtheorem{datastruct}[theorem]{Data Structure}

\theoremstyle{definition}
\newtheorem{defn}[theorem]{Definition}
\newtheorem{assumption}[theorem]{Assumption}

\newcommand{\alg}[1]{\textsf{#1}}
\newcommand{\boztas}{Bozta{\c s}}
\newcommand{\cpp}{C\nolinebreak\hspace{-.05em}%
    \raisebox{.4ex}{\tiny\textbf{+}}\nolinebreak\hspace{-.10em}%
    \raisebox{.4ex}{\tiny\textbf{+}}}
\newcommand{\den}{\beta}
\newcommand{\dnaa}{A}
\newcommand{\dnac}{C}
\newcommand{\dnag}{G}
\newcommand{\dnat}{T}
\newcommand{\dragon}{\textsf{Dragon}}
\newcommand{\er}{Erd{\H o}s-R{\'e}nyi}
\newcommand{\error}{\epsilon}
\newcommand{\lpair}[2]{(#1,#2)}
\newcommand{\mmap}[3]{{#1}[\,#2,\,#3\,]}
\newcommand{\num}{\alpha}
\newcommand{\pmax}{{p_\mathrm{max}}}
\newcommand{\pmin}{{p_\mathrm{min}}}
\newcommand{\rank}[1]{r_{#1}}
\newcommand{\undef}{\varnothing}
\newcommand{\vlink}{\mbox{\Large $\downarrow$}}

\title{Searching a bitstream in linear time \\
       for the longest substring of any given density}
\author{Benjamin A.~Burton}

\date{June 3, 2010}

\begin{document}

\maketitle

\begin{abstract}
Given an arbitrary bitstream, we consider the problem of finding
the longest substring whose ratio of ones to zeroes equals a given value.
The central result of this paper is an algorithm that solves this problem
in linear time.
The method involves (i)~reformulating the problem as a constrained walk
through a sparse matrix, and then (ii)~developing a data structure for
this sparse matrix that allows us to perform each step of the walk in
amortised constant time.
We also give a linear time algorithm to find the longest substring whose
ratio of ones to zeroes is bounded below by a given value.
Both problems have practical relevance to cryptography and bioinformatics.
\end{abstract}

\section{Introduction} \label{s-intro}

% ---------- Summary of main problems ----------

Consider a bitstream of length $n$, that is, a sequence of bits
$x_1, x_2, \ldots, x_n$ where each $x_i$ is $0$ or $1$.
We define the \emph{density} of this bitstream to be the proportion
of bits that are equal to one (equivalently, $\sum x_i / n$).
The density always lies in the range $[0,1]$:  a stream of zeroes
has density $0$, a stream of ones has density $1$,
and a stream of random bits should have density close to $\frac12$.

In this paper we are interested in the densities of substrings
within a bitstream.  By a \emph{substring}, we mean a continuous
sequence of bits $x_a,x_{a+1},\ldots,x_{b-1},x_b$, beginning at some
arbitrary position $a$ and ending at some arbitrary position $b$.
The \emph{length} of a substring is the number of bits that it contains
(that is, $b-a+1$), and the \emph{density} of the substring is likewise
the proportion of ones that it contains (that is,
$\sum_{i=a}^b x_i / (b-a+1)$).

In particular, we are interested in the following two problems:

\begin{problem}[Fixed density problem] \label{p-fixed}
    Suppose we are given a bitstream $S$ of length $n$ and a fixed ratio
    $\theta \in [0,1]$.  What is the longest substring of $S$ whose
    density is \emph{equal to} $\theta$?
\end{problem}

\begin{problem}[Bounded density problem] \label{p-bounded}
    Suppose we are given a bitstream $S$ of length $n$ and a fixed ratio
    $\theta \in [0,1]$.  What is the longest substring of $S$ whose
    density is \emph{at least} $\theta$?
\end{problem}

For example, suppose we are given the bitstream $S = 010110101100$ of length
$n=12$.
Then the longest substring with density \emph{equal to} $\theta=0.6$
has length ten ($0\underline{\mathbf{1011010110}}0$),
and the longest substring with density \emph{at least} $\theta=0.7$
has length seven ($010\underline{\mathbf{1101011}}00$).
Note that each problem might have many solutions or
no solution at all.

% ---------- Overview of applications ----------

Both of these problems have important applications for cryptography.
Many cryptographic systems are dependent on pseudo-random number
generators (PRNGs), and any unwanted predictability or structure in a
PRNG becomes a potential attack point for the underlying cryptosystem.
For this reason PRNGs are typically subjected to a stringent series of
randomness tests, such as those described in \cite{knuth97-art2} or
\cite{marsaglia85-view}.

{\boztas} et~al.\ have recently designed a new series of randomness tests
based on the densities of substrings \cite{boztas09-density}.
To construct these tests, they
use the {\er} law of large numbers \cite{arratia90-erdosrenyi,erdos70-law}
to compute the limiting distributions for
solutions to the fixed density problem, the bounded density problem
and related problems.
They then compare observed values against these limiting
distributions, and they have identified a possible weakness in the
{\dragon} stream cipher \cite{chen05-dragon} as a result.

Locating substrings with various density properties also
has important applications in bioinformatics.
A sequence of DNA consists of a long string of nucleotides marked
\dnag, \dnac, \dnat\ or \dnaa,
and subsequences with high proportions of \dnag\ and \dnac\ are called
\emph{GC-rich regions}.
GC-richness is correlated with factors such as gene density
\cite{zoubak96-distribution}, gene length \cite{duret95-longgenes},
recombination rates \cite{fullerton01-recombination},
codon usage \cite{sharp95-silence},
and the increasing complexity of organisms
\cite{bernardi00-isochores,hardison91-alphaglobin}.

To identify GC-rich regions we convert a DNA sequence into
a bitstream, where each \dnag\ or \dnac\ becomes a one bit,
and each \dnat\ or \dnaa\ becomes a zero bit.
We then search for high-density substrings in this bitstream,
using techniques such as those discussed here.

Further applications of density problems in the field of bioinformatics
are discussed by Goldwasser et~al.\ \cite{goldwasser05-maxdense} and
Lin et~al.\ \cite{lin02-length}.
In addition, Greenberg \cite{greenberg03-heavydense} signals potential
applications in the field of image processing.

% ---------- Algorithms + outline of paper ----------

The focus of this paper is on finding fast algorithms to solve
Problems~\ref{p-fixed} and~\ref{p-bounded}.  Both problems allow
simple brute-force algorithms that run in $O(n^2)$ time.
For the fixed density problem, {\boztas} et~al.\ improve on this
with their \alg{SkipMisMatch} algorithm \cite{boztas09-density},
which remains $O(n^2)$ in the worst case but has an improved
average-case time complexity of $O(n \log n)$.
We outline their contribution in Section~\ref{s-quad}.

Our first contribution in this paper is a series of simple algorithms that
solve both the fixed and bounded density problems in $O(n \log n)$ time,
even in the worst case.  These algorithms are easy to implement and
effective in practice, and are based upon a central geometric observation.
We cover these log-linear algorithms in Section~\ref{s-log}.

In Section~\ref{s-fixed} we follow with our main result, which is an
algorithm that solves the fixed density problem in $O(n)$ time,
again in the worst case.  Based on one of the previous log-linear algorithms,
this algorithm introduces a specialised data structure that allows us to
process each bit of the bitstream in amortised constant time.
Broadly speaking, we:
\begin{itemize}
    \item express our bitstream as a sequence of steps through a sparse
    matrix, where each step requires a localised search and possible
    insertion into this matrix;
    \item design a specialised data structure that ``compresses'' this
    sparse matrix, so that each localised search and insertion can be
    performed in amortised constant time.
\end{itemize}
The amortised analysis is based on aggregation---in essence we
count the ``interesting'' steps of the algorithm by associating them
with distinct
elements of the bitstream, thereby showing the number of such
steps to be $O(n)$.  Details of the proof are given in
Section~\ref{s-fixed-proof}.

Our final contribution is in Section~\ref{s-bounded}, where we give an
$O(n)$ time algorithm for the bounded density problem.  In contrast to
the fixed density problem, this final algorithm is quite simple,
involving just a handful of linear scans.

To conclude, we measure the practical performance of our algorithms
in Section~\ref{s-performance}.  It is reassuring to find that our linear
algorithms are worth the extra difficulty, consistently outperforming the
other algorithms for large bitstream lengths $n$.

% ---------- Final notes ----------

In related work, several authors have considered problems of finding
\emph{maximal density} substrings in a bitstream subject to a variety of
constraints.  See in particular work by
Lin et~al.\ \cite{lin02-length}, who place a lower bound on the length
of the substring; Goldwasser et~al.\ \cite{goldwasser05-maxdense},
who improve the prior solution and also place both lower and upper
bounds; and Greenberg \cite{greenberg03-heavydense}, who studies a
variant relating to compressed bitstreams.

Hsieh et~al.\ \cite{hsieh08-interval} study a series of more general
problems, where the bitstream is replaced by a sequence of real numbers,
and the density of a substring becomes the average of the
corresponding subsequence.  In addition to developing algorithms, they show
that several such problems---including the fixed density problem---have
a \emph{lower bound} of $\Omega(n \log n)$ time.
Our linear algorithm effectively breaks through this lower bound in the
case where the input sequence consists entirely of zeroes and ones.

Throughout this paper we measure time complexity in ``number of
operations'', where we treat basic arithmetical operations such as
$+$ and $\times$ as constant-time.

\section{Quadratic Algorithms: {\boztas} et al.} \label{s-quad}

In this section we outline the prior work of {\boztas} et~al., including
a simple $O(n^2)$ brute force algorithm as well as their \alg{SkipMisMatch}
algorithm, which remains $O(n^2)$ in the worst case but becomes
$O(n \log n)$ in the average case.

\begin{assumption} \label{as-theta}
    Throughout this paper we assume that the ratio $\theta$ is given as
    a rational $\theta = \num/\den$, where $\num$ and $\den$ are integers
    in the range $0 \leq \num \leq \den \leq n$, and where
    $\gcd(\num,\den) = 1$.
\end{assumption}

This assumption is not restrictive in any way.  If $\theta$ cannot be
expressed as above then the fixed density problem has no
solution, and for the bounded density problem we can harmlessly
replace $\theta$ with a nearby rational that satisfies
our requirements.

A na\"ive brute force solution runs in $O(n^3)$ time: for each possible
start point and end point, walk through the substring and count the ones.
However, there are several different tricks that can easily convert this
into $O(n^2)$ by replacing ``walk through the substring'' with a
constant time operation.  One such trick is to use a rank table.

\begin{defn}[Rank Table]
    A \emph{rank table} is an array $\rank{0},\rank{1},\ldots,\rank{n}$,
    where each entry $\rank{k}$ counts the number of ones in the
    substring $x_1,\ldots,x_k$.
\end{defn}

In other words, $\rank{k} = \sum_{i=1}^k x_i$.  It is clear that the
complete rank table can be precomputed in $O(n)$ time, and that it
supports constant time queries of the form
``how many ones appear in the substring $x_a,\ldots,x_b$?'' by simply
computing $\rank{b} - \rank{a-1}$.

For the fixed density problem,
the \alg{SkipMisMatch} algorithm further optimises this $O(n^2)$
brute force method by making the following observations:
\begin{enumerate}[(i)]
    \item \label{en-skip-terminate}
    We are searching for the \emph{longest} substring of density
    $\theta$.  We can
    therefore reorganise our search to work from the longest substring down
    to the shortest, allowing us to terminate as soon as we find
    \emph{any} substring of density $\theta$.

    \item \label{en-skip-multiple}
    If we find such a substring, its length must be a multiple of
    $\den$ (where $\theta = \num/\den$ as above).  We can therefore
    restrict our search to substrings of such lengths.

    \item \label{en-skip-error}
    When searching for substrings of length $k\den$, we need to find
    precisely $k\num$ ones to give a density of $\theta$.  If at some point
    we find $k\num \pm \error$ ones, we must step forward
    \emph{at least $\error$ positions}
    in our bitstream before we can ``undo the error'' and
    potentially find the $k\num$ ones that we seek.
\end{enumerate}

\begin{figure}[htb]
    \centering
    \setlength{\fboxsep}{0.5\baselineskip} % The default margin is *very* thin.
    \framebox{\begin{minipage}{0.9\textwidth}\small%
    \begin{algorithmic}[0]
        \Procedure{SkipMisMatch}{$x_1,\ldots,x_n,\ \theta=\num/\den$}
        \State Build a rank table $\rank{0},\rank{1},\ldots,\rank{n}$
        \For{$k \gets \lfloor \frac{n}{\den} \rfloor$
                \textbf{downto}\ $1$}
                \Comment{Search for substrings of length $k\den$}
            \State $(a,b) \gets (1,k\den)$
                \Comment{Initial start and end for our substring}
            \While{$b \leq n$}
                \State $\error \gets |k\num - (\rank{b} - \rank{a-1})|$
                    \Comment{Compute the ``error'' for this substring}
                \If{$\error = 0$}
                    \State Output $(a,b)$ and terminate
                \Else
                    \State $(a,b) \gets (a+\error, b+\error)$
                        \Comment{We can safely skip forward $\error$
                        positions}
                \EndIf
            \EndWhile
        \EndFor
        \State Output ``no such substring'' and terminate
        \EndProcedure
    \end{algorithmic}
    \end{minipage}}
    \caption{The \alg{SkipMisMatch} algorithm for the fixed density problem}
    \label{fig-skipmismatch}
\end{figure}

Bundling these observations together, we obtain the \alg{SkipMisMatch}
algorithm as illustrated in Figure~\ref{fig-skipmismatch}.
The worst-case complexity is clearly still $O(n^2)$, but for a random
bitstream the \emph{expected} performance can be significantly better.
In particular, {\boztas} et~al.\ prove the following result
as a part of \cite[Lemma~4]{boztas09-density}:

\begin{lemma} \label{l-skipmismatch-expected}
    Suppose we have a random bitstream, where each bit is one with
    probability $\rho$ or zero with probability $1-\rho$.
    Then \alg{SkipMisMatch} has expected time bounded by
    \[
        O\left(|\theta - \rho|^{-1} \times
               \frac{n}{\den} \log \frac{n}{\den}\right).
    \]
\end{lemma}

For fixed $\theta$ and $\rho$, this reduces to an expected time of
$O(n \log n)$, as long as $\theta \neq \rho$.  However, if we
retain the dependency on $\theta$ (and hence its denominator $\den$),
we find that \alg{SkipMisMatch} is rewarded by large denominators
$\den$ (which enhance the power of optimisation~(\ref{en-skip-multiple})),
and is penalised by values of $\theta$ close to $\rho$ (which limit the
use of optimisation~(\ref{en-skip-error})).

To summarise, the \alg{SkipMisMatch} algorithm is easy to code and runs
significantly faster than brute force, but its performance depends
heavily on the given value of $\theta$.  In addition, some broader
issues might arise---the expected $O(n \log n)$ time is appropriate
for random bitstreams (as found in cryptographic
applications, for instance), but might not hold for applications such as
bioinformatics and image processing where bitstreams become more
structured.  Moreover, the algorithm does not translate well to the
bounded density problem.  All of these reasons highlight the need for
faster and more robust algorithms, which form the subject of the
remainder of this paper.

\section{Log-Linear Algorithms: Maps and Sorting} \label{s-log}

In this section we introduce our first truly sub-quadratic
algorithms for solving the fixed and bounded density problems.
We describe \alg{DistMap}, a simple algorithm involving a map
structure, and \alg{DistSort}, a variation that
replaces this map with a sort and a linear scan.  Both of these algorithms
run in $O(n \log n)$ time, even in the worst case.

Although we present even faster algorithms in Sections~\ref{s-fixed}
and~\ref{s-bounded}, both \alg{DistMap} and \alg{DistSort}
are simple to describe and easy to implement.  Moreover, both algorithms
play important roles:  \alg{DistMap} is the foundation upon which the
linear algorithm of Section~\ref{s-fixed} is built, and
\alg{DistSort} is a more flexible variant that can solve both
the fixed and bounded density problems.

\subsection{Graphical Representations}

Our first step in developing these sub-quadratic algorithms
is to find a graphical representation for our bitstreams.

\begin{figure}[htb]
    \centering
    \includegraphics{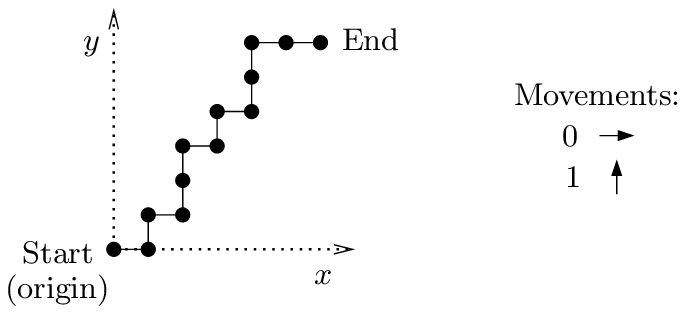}
    \caption{The grid representation for the bitstream $010110101100$}
    \label{fig-walk}
\end{figure}

\begin{defn}[Grid Representation]
    We can plot any
    bitstream as a walk through an infinite two-dimensional grid as
    follows.\footnote{This is related to, but not the same as,
    the walk through the sparse matrix that we use for the linear algorithm
    in Section~\ref{s-fixed}.}
    We begin at the origin $(0,0)$, and then step one unit in the
    $x$-direction each time we encounter a zero, or one unit in the
    $y$-direction each time we encounter a one, as illustrated in
    Figure~\ref{fig-walk}.
    We refer to this as the \emph{grid representation} of the bitstream.
\end{defn}

All of the new algorithms developed in this paper are based upon the
following simple geometric observation:

\begin{lemma} \label{l-slope}
    A substring of a bitstream has density $\theta$ if and only if
    the line joining its start and end points in the grid representation
    has gradient $\frac{\theta}{1 - \theta}$.
\end{lemma}

\begin{figure}[htb]
    \centering
    \includegraphics{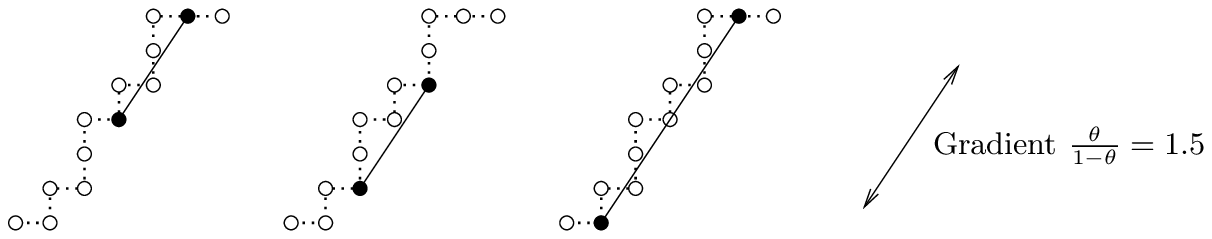}
    \caption{Pairs of points that represent
        substrings of density $\theta = 0.6$}
    \label{fig-slopes}
\end{figure}

To illustrate, Figure~\ref{fig-slopes} builds on the previous example by
searching for substrings of density $\theta = 0.6$.  Several pairs of
points separated by gradient $\frac{\theta}{1-\theta} = 1.5$ are marked
(though there are several more such pairs that are not marked).
The first two pairs correspond to substrings of length five, and the
third pair corresponds to a substring of length ten.

\begin{figure}[htb]
    \centering
    \includegraphics{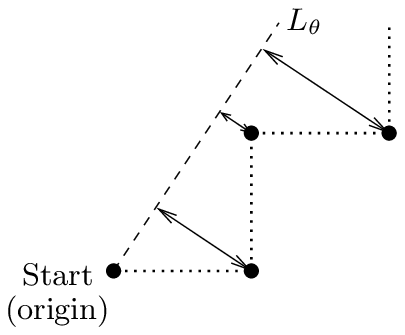}
    \caption{Measuring the distance of each point from the line $L_\theta$}
    \label{fig-dist}
\end{figure}

We can find such pairs of points by drawing a line $L_\theta$
through the origin with slope $\frac{\theta}{1-\theta}$, and then
measuring the \emph{distance} of each point from this line
(where distances are signed, so that points above or below the line
have positive or negative distance respectively).
This is illustrated in Figure~\ref{fig-dist}.
It is clear that two points are joined by a line of gradient
$\frac{\theta}{1-\theta}$ if and only if their distances from
$L_\theta$ are the same.

Although such distances can be messy to compute, with appropriate
rescaling we can convert them into integers as follows.

\begin{defn}[Distance Sequence] \label{d-dist}
    Recall from Assumption~\ref{as-theta} that $\theta = \num/\den$,
    where $\gcd(\num,\den) = 1$.  For a given bitstream $x_1,\ldots,x_n$,
    we define the \emph{distance sequence} $d_0,d_1,\ldots,d_n$ by
    the formula
    \[ d_i = (\den-\num) \cdot (\mbox{number of ones in $x_1,\ldots,x_i$})
    - \num \cdot (\mbox{number of zeroes in $x_1,\ldots,x_i$}). \]
    In other words,
    $d_i = (\den-\num) \rank{i} - \num (i - \rank{i})
         = \den \rank{i} - \num i$,
    where $\rank{i}$ is the corresponding entry in the rank table.
\end{defn}

With a little thought it can be seen that $d_i$ is proportional to the
distance from $L_\theta$ of the point at the end of the $i$th step of the
walk.  This empowers the distance sequence with the following critical
property:

\begin{lemma} \label{l-dist}
    The substring $x_a,\ldots,x_b$ has density equal to $\theta$
    if and only if $d_{a-1} = d_b$.
    Similarly, the substring $x_a,\ldots,x_b$ has density at least $\theta$
    if and only if $d_{a-1} \leq d_b$.
\end{lemma}

\begin{proof}
    Although this follows immediately from the geometric argument
    above, we can also prove it directly.
    Using the formula
    $d_i = \den \rank{i} - \num i$,
    we find that $d_{a-1} = d_b$ if and only if
    $\den(\rank{b}-\rank{a-1}) = \num(b-a+1)$, or equivalently
    \[ \mbox{density of $x_a,\ldots,x_b$} =
        \frac{\rank{b} - \rank{a-1}}{b-a+1} = \frac{\num}{\den} = \theta. \]
    The argument for density $\geq \theta$ is similar.
\end{proof}

\subsection{The \alg{DistMap} Algorithm}

With Lemma~\ref{l-dist} we now have a simple solution to the fixed density
problem.  We compute the distance sequence $d_0,\ldots,d_n$ as we
pass through our bitstream, keeping track of which distances we have
seen before and when we first saw them.  Whenever we find that a distance
\emph{has} been seen before, we have a substring of density
$\theta$ and therefore a potential solution.

We keep track of previously-seen distances using a
$\mathit{key} \mapsto \mathit{value}$ map structure with
worst-case $O(\log n)$ search and insertion, such as a red-black tree
\cite{cormen01-algorithms}.  Here the key is a distance $D$
that we have seen before, and the value is the position at which we
first saw it (i.e., the smallest $i$ for which $d_i = D$).

\begin{figure}[htb]
    \centering
    \setlength{\fboxsep}{0.5\baselineskip} % The default margin is *very* thin.
    \framebox{\begin{minipage}{0.9\textwidth}\small%
    \begin{algorithmic}[0]
        \Procedure{DistMap}{$x_1,\ldots,x_n,\ \theta=\num/\den$}
        \State $(a,b) \gets (0,0)$ \Comment{Best start/end found so far}
        \State $\delta \gets 0$ \Comment{Current distance $d_i$}
        \State Initialise the empty map $m$

        \Statex

        \State Insert $m[0] \gets 0$
            \Comment{Record the starting point $d_0 = 0$}
        \For{$i \gets 1$ \textbf{to}\ $n$}
            \If{$x_i = 1$} \Comment{Compute the new distance $d_i$}
                \State $\delta \gets \delta + (\den-\num)$
            \Else
                \State $\delta \gets \delta - \num$
            \EndIf

            \Statex

            \If{$m$ has no key $\delta$}
                    \Comment{Have we seen this distance before?}
                \State Insert $m[\delta] \gets i$
                    \Comment{No, this is the first time}
            \Else
                \If{$i - m[\delta] > b - a + 1$}
                        \Comment{Yes, back at position $m[\delta]$}
                    \State $(a,b) \gets (m[\delta]+1,i)$
                        \Comment{Longest substring found so far}
                \EndIf
            \EndIf
        \EndFor

        \Statex

        \State Output $(a,b)$
        \EndProcedure
    \end{algorithmic}
    \end{minipage}}
    \caption{The \alg{DistMap} algorithm for the fixed density problem}
    \label{fig-distmap}
\end{figure}

The result is the algorithm \alg{DistMap}, described in
Figure~\ref{fig-distmap}.  Given our choice of map structure, the
following result is clear:

\begin{lemma}
    The algorithm \alg{DistMap} solves the fixed density problem
    in $O(n \log n)$ time in the worst case.
\end{lemma}

We could of course use a hash table instead of a map structure---with a
judicious choice of hash function this could yield $O(n)$ expected time,
though the worst case could potentially be much slower.
Because we offer a worst-case $O(n)$ algorithm in Section~\ref{s-fixed},
we do not pursue hashing any further here.

\subsection{The \alg{DistSort} Algorithm}

We move now to a variant of \alg{DistMap} that removes any
need for a map structure at all.  Instead, we replace this map with a
simple array that we sort in-place after all $n$ bits of the bitstream
have been processed.  The new algorithm is named \alg{DistSort}, and has
the following advantages:
\begin{itemize}
    \item Whilst the map structure plays a key role in giving us
    $O(n \log n)$ running time, it also comes with a non-trivial memory
    overhead.  If $n$ is large and memory becomes a problem,
    the in-place sort used by \alg{DistSort} may be a more economical choice.

    \item \alg{DistMap} relies on searching for precise
    matches $d_{a-1} = d_b$ within the map structure.
    This makes it unsuitable for the \emph{bounded} density problem,
    which requires only $d_{a-1} \leq d_b$ (Lemma~\ref{l-dist}).
    If we replace our map with an array sorted by distance $d_i$, then
    both problems become easy to solve.  Indeed, we find with \alg{DistSort}
    that the solutions for the fixed and bounded density problems
    differ by just one line.
\end{itemize}

\noindent % This short sentence is squeezed between two itemized lists,
          % and the indent makes it harder to spot.
The key ideas behind \alg{DistSort} are as follows:
\begin{itemize}
    \item We walk through the bitstream and compute each distance
    $d_i$ as we go, just as we did for \alg{DistMap}.
    However, instead of storing distances in a map, we
    store each pair $(d_i, i)$ in a simple array $z[0..n]$,
    so that each array entry $z[i]$ is the pair $(d_i,i)$.

    \item Once we have finished our walk through the bitstream,
    we sort the array $z[0..n]$ by distance.  This gives us a sequence
    of $(\mathrm{distance},\,\mathrm{position})$ pairs
    \[ (D_0, P_0)\quad(D_1, P_1)\quad\ldots\quad(D_n, P_n), \]
    where $D_0 \leq D_1 \leq \ldots \leq D_n$ and where each
    $D_i$ is the distance after the $P_i$th step.

    \item Finding positions with matching distances is now a simple
    matter of walking through the array from left to right---all of the
    positions with the same distance will be clumped together.
    In each clump we track the smallest and largest positions
    $\pmin$ and $\pmax$, and these become a candidate substring
    $x_{(\pmin+1)},\ldots,x_\pmax$ with density $\theta$.
    The longest such substring is then our solution to the fixed density
    problem.

    \item Solving the bounded density problem is just as easy.
    The only difference is that we now need our substring
    $x_{(\pmin+1)},\ldots,x_\pmax$ to satisfy $d_\pmin \leq d_\pmax$,
    not $d_\pmin = d_\pmax$.
    To achieve this, we simply change $\pmin$ from the smallest
    position in \emph{this clump} to the smallest position in
    \emph{all clumps seen so far}.
\end{itemize}

\begin{figure}[htb]
    \centering
    \setlength{\fboxsep}{0.5\baselineskip} % The default margin is *very* thin.
    \framebox{\begin{minipage}{0.9\textwidth}\small%
    \begin{algorithmic}[0]
        \Procedure{DistSort}{$x_1,\ldots,x_n,\ \theta=\num/\den$}
        \State Initialise an array $z[0..n]$ of $(\mbox{dist},\mbox{pos})$ pairs
        \State $\delta \gets 0$ \Comment{Current distance $d_i$}

        \Statex

        \State $z[0] \gets (0,0)$
            \Comment{Record the starting point $d_0 = 0$}
        \For{$i \gets 1$ \textbf{to}\ $n$}
            \If{$x_i = 1$} \Comment{Compute the new distance $d_i$}
                \State $\delta \gets \delta + (\den-\num)$
            \Else
                \State $\delta \gets \delta - \num$
            \EndIf

            \State $z[i] \gets (\delta, i)$
                \Comment{Store the pair $(d_i, i)$ in our array}
        \EndFor
        \Statex
        \State Sort $z[0..n]$ by distance, giving a sorted sequence \\
            \hspace{\algorithmicindent}%
            \hspace{\algorithmicindent}%
            of pairs $(D_0, P_0)\;(D_1, P_1)\;\ldots\;(D_n, P_n)$
        \Statex
        \State $(a,b) \gets (0,0)$ \Comment{Best start/end positions
            found so far}
        \State $(\pmin, \pmax) \gets (P_0,P_0)$
            \Comment{Potential start/end positions}
        \State $i \gets 0$
        \While{$i \leq n$}
            \State $\pmin \gets P_i$
                \Comment{\textbf{Do this for the fixed density problem ONLY}}
                \label{line-distsort-fixedonly}
            \State $\pmax \gets P_i$

            \Statex

            \State $i \gets i + 1$
                \Comment{Run through a clump of pairs with the same distance}
            \While{$i \leq n$ and $D_i = D_{i-1}$}
                \If{$P_i < \pmin$}
                    \State $\pmin \gets P_i$
                        \Comment{A smaller position with this distance}
                \EndIf
                \If{$P_i > \pmax$}
                    \State $\pmax \gets P_i$
                        \Comment{A larger position with this distance}
                \EndIf
                \State $i \gets i + 1$
            \EndWhile

            \Statex

            \If{$\pmax - \pmin > b - a + 1$}
                \State $(a,b) \gets (\pmin + 1, \pmax)$
                    \Comment{Longest substring found so far}
            \EndIf
        \EndWhile

        \Statex

        \State Output $(a,b)$
        \EndProcedure
    \end{algorithmic}
    \end{minipage}}
    \caption{The \alg{DistSort} algorithm for the fixed and bounded
        density problems}
    \label{fig-distsort}
\end{figure}

The full algorithm is given in Figure~\ref{fig-distsort}.
The fixed and bounded density algorithms differ by only one line (marked
with a comment in bold), where in the bounded case we do not reset
$\pmin$ upon entering a new clump of pairs with equal distances.

Regarding time complexity,
we can choose a worst-case $O(n \log n)$ sorting algorithm, such as the
\alg{introsort} algorithm of Musser \cite{musser97-introsort}.
The subsequent scan through the array runs in linear time, yielding
the following overall result:

\begin{lemma}
    The algorithm \alg{DistSort} solves both the fixed and bounded
    density problems in $O(n \log n)$ time in the worst case.
\end{lemma}

\section{Solving the Fixed Density Problem} \label{s-fixed}

We proceed now to an algorithm for the fixed density problem that runs
in $O(n)$ time, even in the worst case.  This algorithm uses
\alg{DistMap} as a starting point, but
replaces the generic map structure with a specialised data structure
for the task at hand.

The central observation is the following.  As we run the \alg{DistMap}
algorithm,
\emph{each successive key in our map is always obtained by adding
$+(\den-\num)$ or $-\num$ to the previous key}.  We exploit this
constraint to design a data structure that allows us
to ``jump'' from one key to the next without requiring a full search,
thereby eliminating the $\log n$ factor from our running time.

The data structure is fairly detailed, making it difficult to give
a simple overview.  The following outline summarises the broad ideas
involved, but for a clearer picture the reader is referred to the
full description in Sections~\ref{s-fixed-theory}
and~\ref{s-fixed-imp}.  The running time of $O(n)$ is established in
Section~\ref{s-fixed-proof} using amortised analysis.

\begin{itemize}
    \item We begin by arranging the integers into an infinite
    two-dimensional lattice (Figure~\ref{fig-lattice-intro}),
    so that $+(\den-\num)$ represents a single step to the right
    and $-\num$ represents a single step down.  This makes
    moving from one key to the next a \emph{local movement}
    within the lattice.
    This lattice has infinitely many columns but only $\beta-\alpha$ rows,
    so a step down from the bottom row wraps back around to the top
    (but with a shift).

    \begin{figure}[htb]
        \centering
        \includegraphics[scale=0.8]{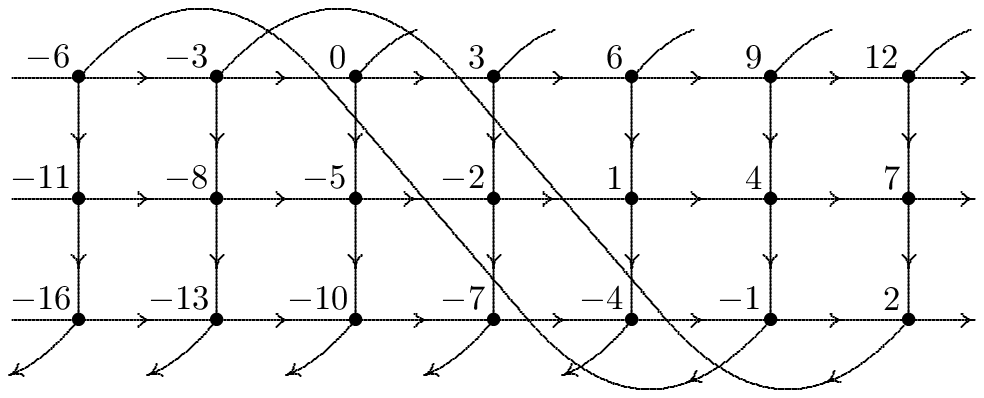}
        \caption{The two-dimensional lattice of integers for $\den-\num=3$
            and $\num=5$.}
        \label{fig-lattice-intro}
    \end{figure}

    \item We now use this integer lattice as the ``domain''
    of our map, so that keys (the distances $d_i$) become points in the
    lattice, and values (the corresponding positions $i$)
    are stored at these points.
    In this way \emph{our data structure becomes a matrix},
    which is sparse because only $n$ points in the lattice
    correspond to ``real'' keys with non-empty values.

    \item The next stage in our design is to ``compress'' this sparse matrix
    by storing not individual $\mathit{key} \mapsto \mathit{value}$ pairs
    but rather \emph{horizontal runs of consecutive pairs}, as illustrated
    in Figure~\ref{fig-compress-intro}.  Storing just the start and end
    of each run allows us to completely reconstruct the
    missing keys and values in between.

    \begin{figure}[htb]
        \centering
        \includegraphics[scale=0.8]{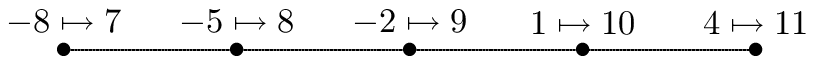}
        \qquad
        {\Large $\mathbf{\Longleftrightarrow}$}
        \qquad
        \includegraphics[scale=0.8]{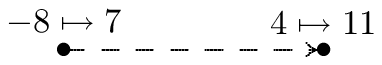}
        \caption{Compressing a horizontal run of consecutive pairs}
        \label{fig-compress-intro}
    \end{figure}

    \item We finish by developing a linked structure for storing our matrix.
    The compressed runs in each row are stored as a ``horizontal''
    linked list, with additional ``vertical'' links between rows
    for downward steps.  We also chain vertical
    links together, yielding a perfect balance that offers
    enough information to support fast movement between keys,
    but enough flexibility to support fast insertion of new
    $\mathit{key} \mapsto \mathit{value}$ pairs.
\end{itemize}

Before presenting the details, it becomes useful to strengthen our
base assumptions as follows.

\begin{assumption} \label{as-nontrivial}
    Recall from Assumption~\ref{as-theta} that
    $\theta=\num/\den$, where $0 \leq \num \leq \den \leq n$.
    From here onwards we strengthen this by assuming the
    stricter bounds $0 < \num < \den \leq n$.
    In other words, we explicitly disallow the special cases
    $\theta = 0$ and $\theta = 1$.
\end{assumption}

Like our earlier assumptions, this is not restrictive in
any way.  If $\theta=0$ or $\theta=1$ then we simply require
the longest continuous substring of zeroes or ones, which is trivial to
find in linear time.

\subsection{The Mapping Matrix} \label{s-fixed-theory}

We begin the details with a formal definition of the integer lattice
depicted in Figure~\ref{fig-lattice-intro}.  Recall from
Assumptions~\ref{as-theta} and~\ref{as-nontrivial} that both
$\den-\num$ and $\num$ are strictly positive,
and that $\gcd(\den-\num,\num) = 1$.

\begin{defn}[Lattice Coordinates] \label{d-coords}
    Let $z$ be any integer.  The \emph{lattice coordinates} of
    $z$ are the unique solutions $\lpair{r}{c}$ to the
    equation
    \begin{equation} \label{eqn-coords}
    (\den-\num) c - \num r = z,
    \end{equation}
    for which $r$ and $c$ are integers and
    $0 \leq r < \den-\num$.  We call $r$ and $c$ the
    \emph{row} and \emph{column} of $z$ respectively.
\end{defn}

For example, consider Figure~\ref{fig-lattice-intro} in which
$\den-\num = 3$ and $\num=5$.  The following table lists the lattice
coordinates of several integers $z$:
\[ \begin{array}{l|c|c|c|c|c|c}
   \mbox{Integer $z$} & -3 & 0 & 3 & 6 & 1 & -4 \\
   \hline
   \mbox{Lattice coordinates of $z$} &
   \lpair{0}{-1} & \lpair{0}{0} & \lpair{0}{1} &
   \lpair{0}{2} & \lpair{1}{2} & \lpair{2}{2}
   \end{array} \]
These are precisely the locations at which each integer can be found
in Figure~\ref{fig-lattice-intro}, where we number the rows and
columns so that the integer zero appears at coordinates $\lpair{0}{0}$.

With a little modular arithmetic it can shown that every integer appears
once and only once in our lattice, as expressed formally
by the following result.
The proof is elementary, and we do not repeat it here.

\begin{lemma} \label{l-coords}
    Lattice coordinates are always well-defined, that is,
    equation~(\ref{eqn-coords}) has a unique solution for every integer $z$.
    Moreover, every pair of integers $\lpair{r}{c}$ with
    $0 \leq r < \den-\num$ forms the lattice coordinates of one and
    only one integer.
\end{lemma}

It is worth reiterating a key feature of this construction, which is that
each bit of the bitstream gives rise to a \emph{local movement}
within the lattice:

\begin{lemma} \label{l-step}
    Consider some position $i$ within the bitstream, where
    $0 \leq i < n$.  Suppose that the lattice coordinates of
    the distance $d_i$ are $\lpair{r}{c}$.  Then:
    \begin{itemize}
        \item If the $(i+1)$th bit is a one, the lattice coordinates
        of the subsequent distance $d_{i+1}$ are
        $\lpair{r}{c+1}$.  That is, we take one step to the right.
        \item If the $(i+1)$th bit is a zero and $r < \den-\num-1$,
        then the lattice coordinates of $d_{i+1}$ are $\lpair{r+1}{c}$.
        That is, we take one step down.
        \item If the $(i+1)$th bit is a zero and $r = \den-\num-1$
        (i.e., we are on the bottom row of the lattice), then
        the lattice coordinates of $d_{i+1}$ are $\lpair{0}{c-\num}$.
        That is, we wrap back around to the top with a shift
        of $\num$ columns to the left.
    \end{itemize}
\end{lemma}

This is a straightforward consequence of Definitions~\ref{d-dist}
and~\ref{d-coords}, and again we omit the proof.
The various movements described in this result are indicated by
the solid lines in Figure~\ref{fig-lattice-intro}.

Recall that our overall strategy is to build a replacement data structure
for the generic $\mathit{key} \mapsto \mathit{value}$ map, whose keys
are distances $d_i$ and whose values are the corresponding
positions $i$ in the bitstream.  Using Lemma~\ref{l-coords} we can replace
each distance $d_i$ with its \emph{lattice coordinates} $(r,c)$,
thereby replacing the old mapping $d_i \mapsto i$ with the new mapping
$(r,c) \mapsto i$.
This effectively gives us a matrix with $\den-\num$ rows and
infinitely many columns, which we formalise as follows.

\begin{defn}[Mapping Matrix]
    We define the \emph{mapping matrix} to be an infinite matrix with
    precisely $\den-\num$ rows (numbered $0,\ldots,\den-\num-1$) and
    infinitely many columns in both directions
    (numbered $\ldots,-1,0,1,\ldots$).  Each cell of this matrix may
    contain an integer, or may contain the symbol $\undef$ representing
    an \emph{empty cell}.  The entry in row $r$ and column $c$
    of the mapping matrix $M$ is denoted $\mmap{M}{r}{c}$.
\end{defn}

Our algorithm now runs as follows.  As we process each bit of the
bitstream, we walk through the cells of the mapping matrix as described
by Lemma~\ref{l-step}.  If we step into an empty cell, we store the
current position in the bitstream.  If we step into a
previously-occupied cell
then we have found a substring of density $\theta$.

\begin{figure}[htb]
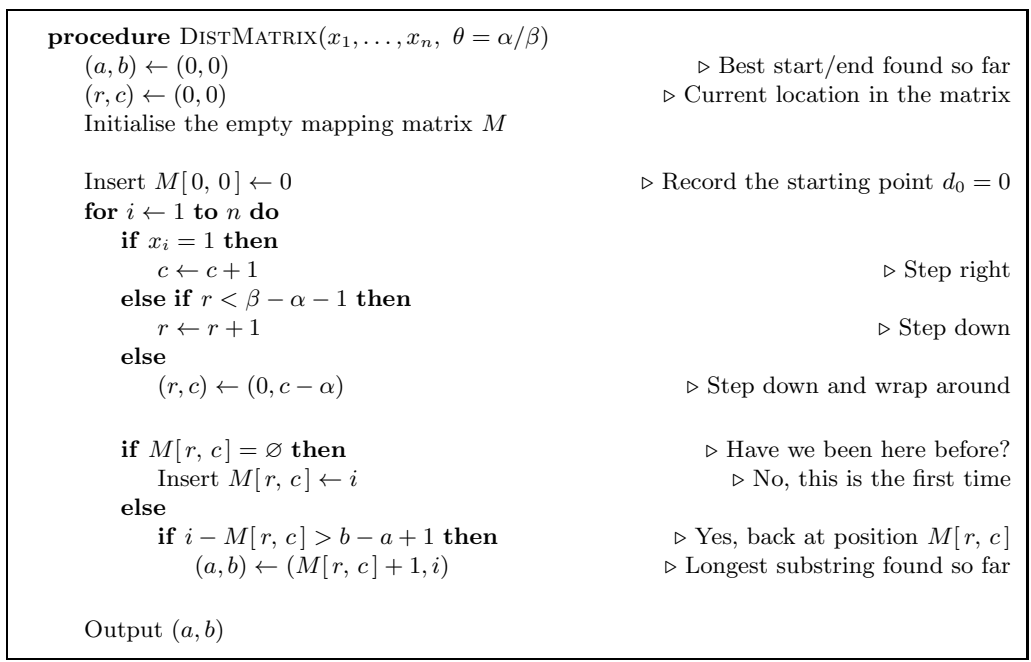

    \centering
    \setlength{\fboxsep}{0.5\baselineskip} % The default margin is *very* thin.
    \framebox{\begin{minipage}{0.9\textwidth}\small%
    \begin{algorithmic}[0]
        \Procedure{DistMatrix}{$x_1,\ldots,x_n,\ \theta=\num/\den$}
        \State $(a,b) \gets (0,0)$ \Comment{Best start/end found so far}
        \State $(r,c) \gets (0,0)$ \Comment{Current location in the matrix}
        \State Initialise the empty mapping matrix $M$

        \Statex

        \State Insert $\mmap{M}{0}{0} \gets 0$
            \Comment{Record the starting point $d_0 = 0$}
        \For{$i \gets 1$ \textbf{to}\ $n$}
            \If{$x_i = 1$}
                \State $c \gets c + 1$ \Comment{Step right}
            \ElsIf{$r < \den-\num-1$}
                \State $r \gets r + 1$ \Comment{Step down}
            \Else
                \State $(r,c) \gets (0,c-\num)$
                    \Comment{Step down and wrap around}
            \EndIf

            \Statex

            \If{$\mmap{M}{r}{c} = \undef$}
                    \Comment{Have we been here before?}
                \State Insert $\mmap{M}{r}{c} \gets i$
                    \Comment{No, this is the first time}
            \Else
                \If{$i - \mmap{M}{r}{c} > b - a + 1$}
                        \Comment{Yes, back at position $\mmap{M}{r}{c}$}
                    \State $(a,b) \gets (\mmap{M}{r}{c}+1,i)$
                        \Comment{Longest substring found so far}
                \EndIf
            \EndIf
        \EndFor

        \Statex

        \State Output $(a,b)$
        \EndProcedure
    \end{algorithmic}
    \end{minipage}}
    \caption{The \alg{DistMatrix} algorithm for the fixed density problem}
    \label{fig-distmatrix}
\end{figure}

The full pseudocode is given in Figure~\ref{fig-distmatrix}, under the
algorithm name \alg{DistMatrix}.  The algorithm is of course
remarkably similar to \alg{DistMap} (Figure~\ref{fig-distmap}), since
the key difference is in the underlying data structure.
Our focus in Section~\ref{s-fixed-imp} is now to fully describe this
data structure, and thereby describe the critical tasks of
evaluating and setting the matrix entry $\mmap{M}{r}{c}$.

\subsection{The Data Structure} \label{s-fixed-imp}

We cannot afford to store the mapping matrix as a two-dimensional array,
because---even ignoring the infinitely many columns---there are $O(n^2)$
\emph{potential} cells that a bitstream of length $n$ might
reach.\footnote{This of course depends upon the value of $\theta$.  If
$\theta=\frac12$ for instance, then there are only $2n+1$ potential cells
and a more direct linear algorithm becomes possible.  Here we treat
the general case $0 < \num < \den \leq n$.}
However, only $n+1$ cells are visited (and hence non-empty)
for any \emph{particular} input bitstream.
That is, \emph{the mapping matrix is sparse}.

We therefore aim for a linked structure, where only the cells we visit
are stored in memory, and where these cells include pointers to nearby
cells to assist with navigation around the matrix.

However, before describing this linked structure we introduce a form of
compression, where we only need to store the cells
involved in \emph{downward steps}.  As we will see in
Section~\ref{s-fixed-proof}, this compression is critical for
stepping through the matrix in amortised constant time.

Our compression relies on the observation that a run of $k$ consecutive
steps to the right produces a sequence of $k$ consecutive values
in the matrix:
\[ \begin{array}{c|c|c|c|c|c}
   \cline{1-6}
   ~ & i & i+1 & \cdots & i+k & ~ \\
   \cline{1-6}
\end{array} \]
We can describe such a sequence by storing only the
start and end points, without having to store
each individual cell in between.

\begin{figure}[tb] % Disallow [h] since the figure is big and complex.
    \centering
    \subfigure[Several paths that cross through a single matrix row]{%
    \label{sub-compress-paths}%
        \includegraphics{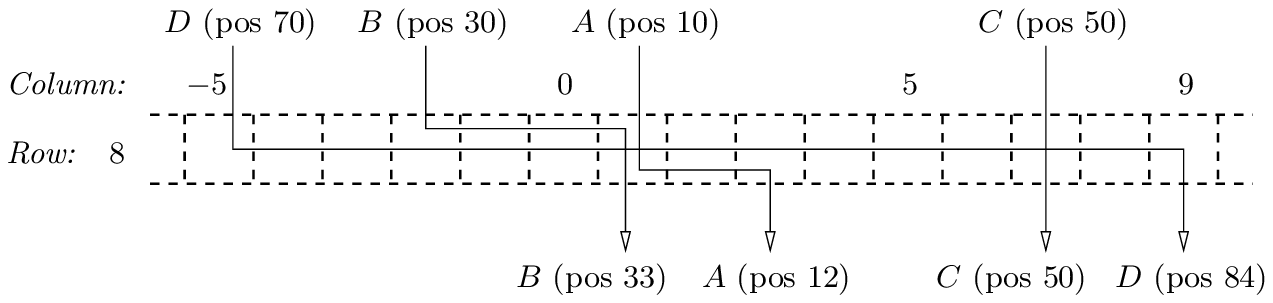}}\\
    \subfigure[The corresponding values in the mapping matrix]{%
    \label{sub-compress-values}%
    \small $\begin{array}{l@{\quad}c|c|c|c|c|c|c|c|c|c|c|c|c|c|c|c|c}
       \textit{Column:} &
       \multicolumn{1}{c}{~} & \multicolumn{1}{c}{-5} &
       \multicolumn{4}{c}{~} & \multicolumn{1}{c}{0} &
       \multicolumn{4}{c}{~} & \multicolumn{1}{c}{5} &
       \multicolumn{3}{c}{~} & \multicolumn{1}{c}{9} \\
       \multicolumn{2}{c}{~} & \multicolumn{1}{c}{\vlink} &
       \multicolumn{2}{c}{~} & \multicolumn{1}{c}{\vlink} &
       \multicolumn{2}{c}{~} & \multicolumn{1}{c}{\vlink} &
       \multicolumn{5}{c}{~} & \multicolumn{1}{c}{\vlink} \\
       \cline{2-18}
       \textit{Row:}\quad 8 & ~ &
         70 & 71 & 72 & 30 & 31 & 32 & 10 & 11 & 12 & 79 & 80 & 81 &
         50 & 83 & 84 & ~ \\
       \cline{2-18}
       \multicolumn{8}{c}{~} & \multicolumn{1}{c}{\vlink} &
       \multicolumn{1}{c}{~} & \multicolumn{1}{c}{\vlink} &
       \multicolumn{3}{c}{~} & \multicolumn{1}{c}{\vlink} &
       \multicolumn{1}{c}{~} & \multicolumn{1}{c}{\vlink}
    \end{array}$}\\
    \subfigure[Storing these values in memory]{%
    \label{sub-compress-storage}%
    \small \begin{tabular}{c|c|c}
        \textit{Cell} & \textit{Value in this cell} &
        \textit{Value to start this run} \\
        \hline
        $(8,-5)$ & 70 & 70 \\
        $(8,-2)$ & 30 & 30 \\
        $(8,1)$ &  10 & 10 \\
        $(8,3)$ &  12 & 78 \\
        $(8,7)$ &  50 & 82 \\
        $(8,9)$ &  84 & $\undef$
    \end{tabular}}
    \caption{Compressing a row of the mapping matrix}
    \label{fig-compress-detail}
\end{figure}

This pattern becomes more complicated when new paths through the
matrix cross over old paths, but the core idea remains the same---we look for
horizontal runs of consecutive values in the matrix, and record only
where they start and end.  Figure~\ref{fig-compress-detail} gives an
example, where four different paths from four different sections of
the bitstream cross through the same row of the matrix.
\begin{itemize}
    \item Figure~\ref{sub-compress-paths} shows the four paths, which
    are labelled $A$, $B$, $C$ and $D$ in chronological order as
    they appear in the bitstream.  For instance, path $A$ enters the row
    at cell $(8,1)$ and position $10$ in the bitstream, takes two steps
    to the right, and exits the row from cell $(8,3)$ at position $12$ in
    the bitstream.  Note that path $B$ subsequently exits from the same cell
    that $A$ entered, and that path $C$ includes no rightward steps at all.

    \item Figure~\ref{sub-compress-values} shows the state of the
    mapping matrix after all four paths have been followed.
    Note that values from older paths take precedence over values from
    newer paths, since we always record the \emph{first} position at
    which we enter each cell.  Vertical arrows are included as reminders
    of the cells at which paths enter and exit the row.

    \item Figure~\ref{sub-compress-storage} shows how this state can be
    ``compressed'' in memory.  We \emph{only store cells at which paths
    enter and exit the row}, and for each such cell $(r,c)$ we record the
    following information:
    \begin{itemize}
        \item The value stored directly in that cell, i.e., $\mmap{M}{r}{c}$;
        \item The value that ``begins'' the horizontal run to the right,
        i.e., $\mmap{M}{r}{c+1} - 1$.
    \end{itemize}
    If the cell $(r,c)$ is itself part of the run (such as $(8,-5)$,
    $(8,-2)$ and $(8,1)$ in our example) then both values will be equal.
    If the cell $(r,c)$ is the exit for an older path (such as
    $(8,3)$ or $(8,7)$ in our example) then these values will be different.
    If there is no run to the right (as with $(8,9)$ in our example)
    then we store the symbol $\undef$.
\end{itemize}

\begin{figure}[htb]
    \centering
    \includegraphics{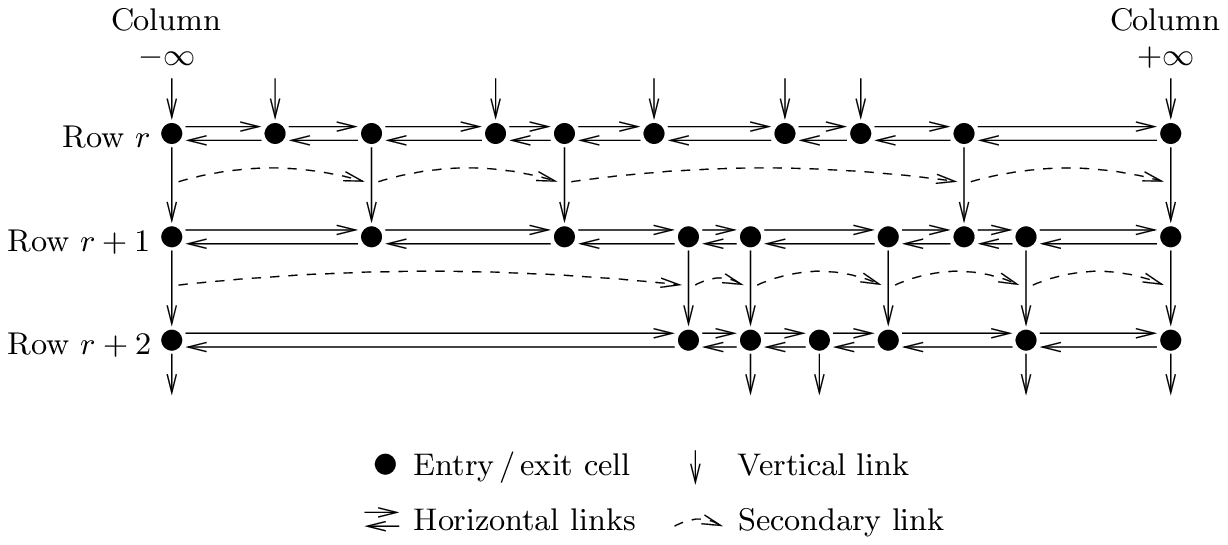}
    \caption{An illustration of the full linked data structure}
    \label{fig-links}
\end{figure}

\noindent % Remove the indent, which otherwise makes this short sentence
          % hard to spot after the itemised list.
We collate this information into a full linked data structure as
described below in Data Structure~\ref{ds-matrix}.
A detailed example of this linked structure is illustrated in
Figure~\ref{fig-links}.

\begin{datastruct}[Mapping Matrix] \label{ds-matrix}
    Suppose we have processed the first $k$ bits of our bitstream.
    To store the current state of the mapping matrix, we keep records
    in memory for the following cells:
    \begin{itemize}
        \item The entry and exit cells in each row, i.e., cells that
        correspond to positions immediately before or after a zero bit;
        \item The two cells corresponding to the beginning of the
        bitstream and our current position;
        \item ``Sentinel'' cells $(r,-\infty)$ and $(r,+\infty)$
        in each row.
    \end{itemize}
    The record for each such cell $(r,c)$ contains the following information:
    \begin{itemize}
        \item The column $c$;
        \item The values $\mmap{M}{r}{c}$ and $\mmap{M}{r}{c+1} - 1$
        as described above, where for the sentinels
        $(r,\pm\infty)$ these values are $\undef$;
        \item Links to the previous and next cells in the same row
        (called \emph{horizontal links}).
    \end{itemize}
    In addition, if we have previously stepped down from this cell then
    we also store:
    \begin{itemize}
        \item A link to the endpoint of this step in the following
        row (called a \emph{vertical link}), where this endpoint is
        $(r+1,c)$ or $(0,c-\num)$ according to whether or not
        $r < \den-\num-1$;
        \item A link that jumps to the \emph{next} vertical link in this
        row, that is, a link to the nearest cell to the right that also
        stores a vertical link (we call this new link a
        \emph{secondary link}).
    \end{itemize}
    We also insert vertical links between the sentinels at
    $(r,\pm\infty)$, running from each row to the next, and join these
    into the chains of secondary links for each row.
\end{datastruct}

To summarise: (i)~the ``interesting'' cells in each row are stored in a
horizontal doubly-linked list, (ii)~we add vertical links corresponding
to previous steps down, and (iii)~we chain together the vertical links from
each row into a secondary linked list.

We return now to fill in the missing parts of the \alg{DistMatrix}
algorithm (Figure~\ref{fig-distmatrix}), namely the evaluation and
setting of the matrix entry $\mmap{M}{r}{c}$.  This can be done
as follows.
\begin{enumerate}[(i)]
    \item At all times we keep a pointer to the current cell in the
    matrix (which, according to Data Structure~\ref{ds-matrix}, always
    has a record explicitly stored).

    \item \label{en-adjust}
    Each time we step right or down, we adjust the data structure
    to reflect the new bit that has been processed, and we move our
    pointer to reflect the new current cell.

    \item Evaluating and setting $\mmap{M}{r}{c}$ then becomes a simple
    matter of dereferencing our pointer.
\end{enumerate}

The only step that might not run in constant time is~(\ref{en-adjust}),
where we adjust the data structure and move our pointer.  The
precise work involved varies according to which type of step we take.
\begin{itemize}
    \item \emph{Step right (processing a one bit):}
    This is a local operation involving no vertical or secondary links.
    We might need to extend the endpoint of the
    current horizontal run or start a new run from the current cell,
    but these are all simple constant time adjustments involving
    only the immediate left and right horizontal neighbours.

    \item \emph{Step down (processing a zero bit):}
    This is a more complex operation that uses
    all three link types.  Suppose that we begin the step in cell
    $(r,c)$; for convenience we assume that we step down to $(r+1,c)$, but the
    wraparound case $r = \den-\num-1$ is much the same.
    If there is already a vertical link $(r,c) \to (r+1,c)$ then we
    simply follow it.  Otherwise we do the following:
    \begin{enumerate}[(1)]
        \item \label{en-down-find}
        Find where the destination cell $(r+1,c)$
        should be inserted in the horizontal list for row $r+1$ (or
        find the cell itself if it is already explicitly stored).
        We do this by:
        \begin{itemize}
            \item walking back along row $r$ until we find the
            nearest vertical link to the left, which we denote $L_-$;
            \item following the secondary link from $L_-$ to
            the nearest vertical link to the right, which we denote $L_+$;
            \item following the link $L_+$ down to row $r+1$;
            \item walking back along row $r+1$ until we find our
            insertion point.
        \end{itemize}

        \begin{figure}[htb]
            \centering
            \subfigure[The neighbourhood of the source cell $(r,c)$]{%
            \label{fig-stepdownpre}%
            \includegraphics{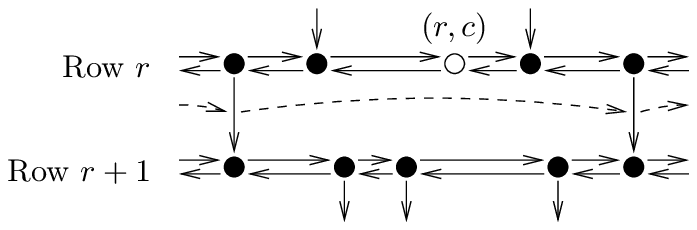}}
            \hspace{1cm}
            \subfigure[The path from $(r,c)$ to $(r+1,c)$]{%
            \label{fig-stepdownpost}%
            \includegraphics{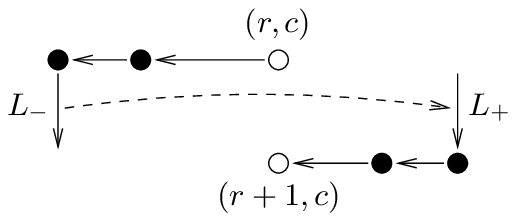}} \\
            \subfigure[The new vertical and secondary links]{%
            \label{fig-stepdownfinal}%
            \includegraphics{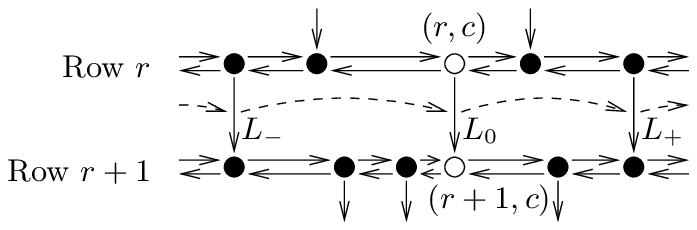}}
            \caption{Stepping down from $(r,c)$ to $(r+1,c)$}
            \label{fig-stepdown}
        \end{figure}

        This series of movements is illustrated in
        Figure~\ref{fig-stepdownpost}.
        Note that our sentinels at $(r,\pm\infty)$ ensure that the
        vertical links $L_-$ and $L_+$ will always exist.
        \item \label{en-down-cell}
        If required, insert the cell $(r+1,c)$ into the horizontal list
        for row $r+1$ and update its immediate horizontal neighbours.
        \item \label{en-down-vertical}
        Insert the new vertical link $(r,c) \to (r+1,c)$, which we
        denote $L_0$.
        \item \label{en-down-secondary}
        Replace the secondary link $L_- \to L_+$ with two secondary
        links $L_- \to L_0 \to L_+$, as illustrated in
        Figure~\ref{fig-stepdownfinal}.
    \end{enumerate}
\end{itemize}

Operations~(\ref{en-down-cell}), (\ref{en-down-vertical})
and~(\ref{en-down-secondary}) are all constant time operations,
but operation~(\ref{en-down-find}) may involve a lengthy walk
through the data structure.  The reason for the convoluted path
(and indeed the secondary links) is because by walking
\emph{backwards} along each row we can ensure
that operation~(\ref{en-down-find}) runs in \emph{amortised} constant
time, as shown in the following section.

\subsection{Analysis of Running Time} \label{s-fixed-proof}

Through the discussions of the previous section, we find that---with the
single exception of the walk from $(r,c)$ to $(r+1,c)$ when we step down
in the mapping matrix---each bit of the bitstream can be processed in
constant time.
The following lemma shows that these exceptional walks can be processed in
\emph{amortised} constant time, giving \alg{DistMatrix}
an overall running time of $O(n)$.

As in the previous section, we assume that we step down from
$(r,c)$ to $(r+1,c)$; the arguments for the wraparound case
$r = \den-\num-1$ are essentially the same.
It is also important to remember that the phrases
\emph{step down} and \emph{step right} refer to the full movement
when processing
some bit of the bitstream, and not the many different links that we might
follow through the data structure in performing such a step.

\begin{lemma} \label{l-amortised}
    Consider the walk from cell $(r,c)$ to $(r+1,c)$ in the ``step down''
    phase of the \alg{DistMatrix} algorithm, as illustrated in
    Figure~\ref{fig-stepdownpost}, and define the \emph{length} of this
    walk to be the total number of links that we follow.
    After processing the entire bitstream, the sum of the
    lengths of all ``step down'' walks is $O(n)$.
    In other words, each such walk can be followed in amortised constant time.
\end{lemma}

\begin{proof}
    We prove this result using aggregate analysis, by ``counting'' the
    number of links in each walk using a rough upper bound.
    The following links are excluded from this count:
    \begin{itemize}
        \item all vertical and secondary links;
        \item the leftmost horizontal link on each row of each walk;
        \item any horizontal links that end at the starting point $(0,0)$;
        \item any horizontal links that end at the current cell $(r,c)$.
    \end{itemize}
    Figure~\ref{fig-longwalkignore} shows a sample walk where the excluded
    links are marked with dotted arrows, and the remaining links
    (all horizontal) are marked with bold solid arrows.
    It is clear that we exclude $O(n)$ links in
    total,\footnote{A horizontal link ending at $(0,0)$ can occur at most
    twice per walk (and at most once if $\den-\num > 1$).
    A horizontal link ending at $(r,c)$ can occur at most once per walk,
    and only in the special case $\den-\num=1$.}
    and so if we can
    show that at most $O(n)$ horizontal links remain then the
    proof is complete.

    \begin{figure}[htb]
        \centering
        \includegraphics{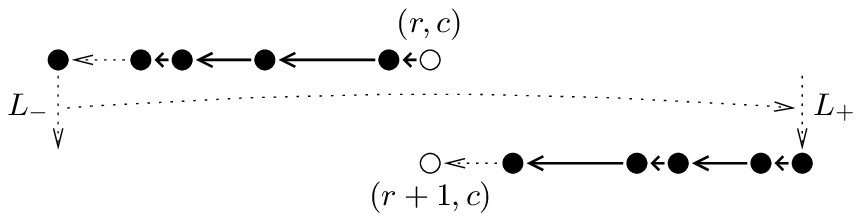}
        \caption{Excluded links in a ``step down'' walk}
        \label{fig-longwalkignore}
    \end{figure}

    Within each walk from $(r,c)$ to $(r+1,c)$, the horizontal links
    that remain have the following critical properties:
    \begin{itemize}
        \item The endpoint of each link in row $r$ is also the endpoint
        of some earlier step down.  Moreover, this earlier step down was
        followed immediately by a succession of steps right that reached
        at least as far along the row as $(r,c)$.

        \item The endpoint of each link in row $r+1$ is also the
        beginning of some earlier step down.  Moreover, this earlier
        step down was preceded immediately by a succession of steps
        right that originated at least as far back along the row as
        $(r+1,c)$.
    \end{itemize}

    \begin{figure}[htb]
        \centering
        \includegraphics{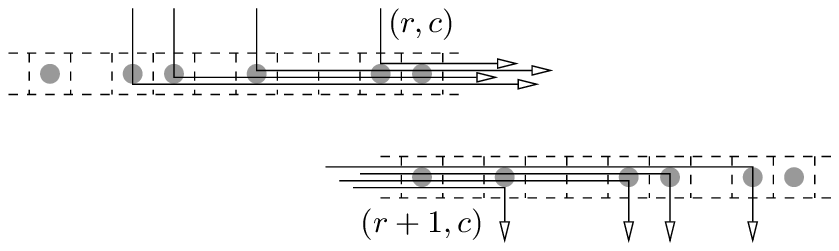}
        \caption{Earlier successions of steps associated
            with the remaining links}
        \label{fig-longwalkright}
    \end{figure}

    These properties are a consequence of our compression (recall that each
    non-sentinel cell that we store is either $(0,0)$, the current cell,
    a row entry or a row exit), as well as the fact that there are no
    vertical links between $L_-$ and $L_+$ that join row $r$ with row
    $r+1$.  Figure~\ref{fig-longwalkright} illustrates the
    successions of rightward steps that are described above.

    We can now associate each remaining link $\ell$ with a position
    $\pi(\ell)$ in the bitstream:
    \begin{itemize}
        \item If the link $\ell$ is on the ``upper'' row $r$, consider
        the oldest sequence of steps that stepped \emph{down} to the endpoint
        of $\ell$ and then \emph{right} all the way across to $(r,c)$, as
        illustrated in Figure~\ref{fig-linkmapupper}.  We define
        $\pi(\ell)$ to be the position in the bitstream that was reached
        by this sequence when it passed through the cell $(r,c)$.
        Note that $0 < \pi(\ell) \leq n$.

        \item If the link $\ell$ is on the ``lower'' row $r+1$, consider the
        oldest sequence of steps that stepped \emph{right} from
        $(r+1,c)$ all the way across to the endpoint of $\ell$ and then
        \emph{down}, as illustrated in Figure~\ref{fig-linkmaplower}.  We
        define $\pi(\ell)$ to be the position in the bitstream that was
        reached by this sequence when it passed through the cell $(r+1,c)$,
        \emph{negated} so that $-n \leq \pi(\ell) < 0$.
    \end{itemize}

    \begin{figure}[htb]
        \centering
        \subfigure[If $\ell$ is on the upper row]{%
        \label{fig-linkmapupper}%
        \includegraphics{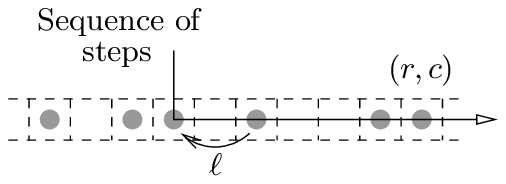}}
        \hspace{1cm}
        \subfigure[If $\ell$ is on the lower row]{%
        \label{fig-linkmaplower}%
        \includegraphics{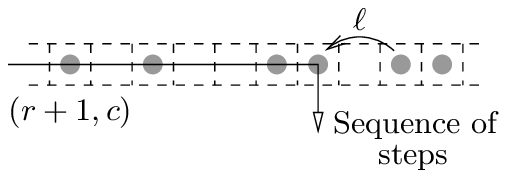}}
        \caption{The earlier sequence of steps that defines $\pi(\ell)$}
        \label{fig-linkmap}
    \end{figure}

    The key to achieving an $O(n)$ total of walk lengths is to
    observe that \emph{the function $\pi$ is one-to-one}:
    \begin{itemize}
        \item A link $\ell_1$ on the upper row of some walk can never
        have the same value of $\pi$ as a link $\ell_2$ on the lower
        row of some (possibly different) walk, since
        $\pi(\ell_2) < 0 < \pi(\ell_1)$.

        \item Within a single walk:
        \begin{itemize}
            \item The values $\pi(\ell)$ for links $\ell$ on the upper row $r$
            are distinct, because each corresponds to a \emph{different
            historical path} through $(r,c)$, with a \emph{different
            initial entry point} into row $r$.

            \item Likewise, the values $\pi(\ell)$ for links $\ell$ on the
            lower row $r+1$ are distinct, because each corresponds to a
            different historical path along row $r+1$ with a different
            final exit point from row $r+1$.
        \end{itemize}

        \item Between different walks:
        \begin{itemize}
            \item Because we insert a new vertical link after every walk,
            each walk must have a distinct starting point $(r,c)$.
            The values $\pi(\ell)$ from the upper rows of different walks
            are therefore distinct because they correspond to
            positions in the bitstream for distinct cells $(r,c)$.

            \item Likewise, the values $\pi(\ell)$ from the lower rows
            of different walks are distinct because they correspond to
            positions in the bitstream for distinct cells $(r+1,c)$.
        \end{itemize}
    \end{itemize}

    Therefore $\pi$ is a one-to-one function.
    Because $\pi(\ell) \in \{-n,-n+1,\ldots,n-1,n\}$,
    it follows that the number of links $\ell$ in the \emph{domain}
    of the function can be at most $2n+1$.  Hence there are
    $O(n)$ horizontal links remaining that we have not
    excluded from our count, and the proof is complete.
\end{proof}

Through Lemma~\ref{l-amortised} we now find that each bit of the bitstream
can be completely processed in amortised constant time,
yielding the following final result:

\begin{corollary}
    The algorithm \alg{DistMatrix} solves the fixed density problem in
    $O(n)$ time in the worst case.
\end{corollary}

\section{Solving the Bounded Density Problem} \label{s-bounded}

We finish our suite of algorithms with a linear time solution to the
bounded density problem, improving upon the log-linear \alg{DistSort}
algorithm of Section~\ref{s-log}.
Unlike our linear time
solution to the fixed density problem, this algorithm is simple
to express, uses no sophisticated data structures, and essentially
involves just a handful of linear scans.

Once again we base our new algorithm on the distance
sequence $d_0,\ldots,d_n$.  Recall from Lemma~\ref{l-dist} that
we seek the longest substring $x_a,\ldots,x_b$ in the bitstream for which
$d_{a-1} \leq d_b$.  We begin with the following simple observation:

\begin{lemma} \label{l-extend}
    Suppose that $x_a,\ldots,x_b$ is the longest substring of
    density $\geq \theta$ in our bitstream.
    Then there is no $i < a-1$ for which $d_i \leq d_{a-1}$,
    and there is no $i > b$ for which $d_i \geq d_b$.
\end{lemma}

The proof is simple---if there were such an $i$, then we could extend
our substring to position $i$ and obtain a longer substring with
density $\geq \theta$.  This result motivates the following definition:

\begin{defn}[Minimal and Maximal Position] \label{d-minmax}
    Let $k$ be a position in the bitstream, i.e.,
    some integer in the range $0 \leq k \leq n$.
    We call $k$ a \emph{minimal position} if there is no
    $i < k$ for which $d_i \leq d_k$, and we call $k$ a
    \emph{maximal position} if there is no $i > k$ for which
    $d_i \geq d_k$.
\end{defn}

\begin{figure}[htb]
    \centering
    \includegraphics[scale=0.9]{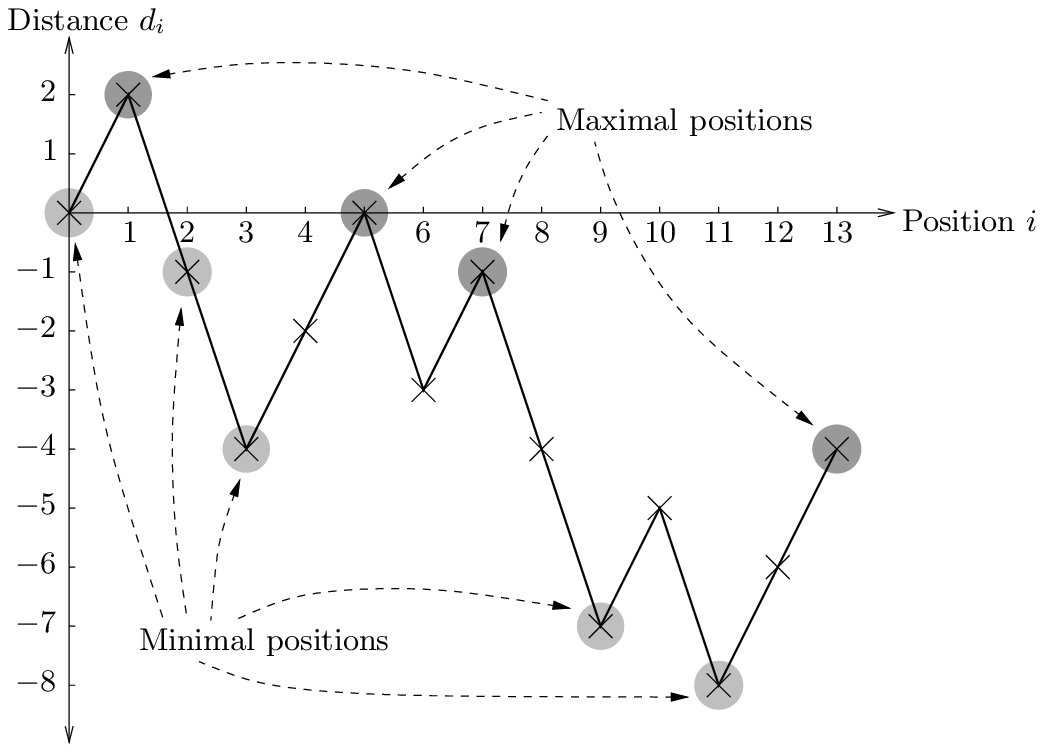}
    \label{fig-minmax}
    \caption{Minimal and maximal positions for the bitstream $1001101001011$}
\end{figure}

Figure~\ref{fig-minmax} plots the distance sequence
for the bitstream $1001101001011$
with target density $\theta = \num/\den = 3/5$,
and marks the minimal and maximal positions on this plot.

Minimal and maximal positions have the following important properties:
\begin{itemize}
    \item \emph{They are the only positions that we need to consider.}
    That is, the solution to the bounded density problem must be a
    substring $x_a,\ldots,x_b$ for which $a-1$ is a minimal position
    and $b$ is a maximal position (Lemma~\ref{l-extend}).

    \item \emph{They are simple to compute in $O(n)$ time.}
    To find all minimal positions, we simply walk through the distance sequence
    $d_0,\ldots,d_n$ and collect positions $i$ for which $d_i$ is
    smaller than any distance seen before.
    To find all maximal positions, we walk through the distance sequence
    in \emph{reverse} ($d_n,\ldots,d_0$)
    and collect positions $i$ for which $d_i$ is
    larger than any distance seen before.

    \item \emph{They are ordered by distance.}
    That is, if the minimal positions are $a_1,a_2,\ldots,a_p$
    from left to right ($a_1 < a_2 < \ldots < a_p$) then we have
    $d_{a_1} > d_{a_2} > \ldots > d_{a_p}$.
    Likewise, if the maximal positions are $b_1,b_2,\ldots,b_q$
    from left to right ($b_1 < b_2 < \ldots < b_q$) then we have
    $d_{b_1} > d_{b_2} > \ldots > d_{b_q}$.
    This is an immediate consequence of Definition~\ref{d-minmax}.
\end{itemize}

\noindent % We have a short sentence squeezed between two indented
          % lists.  If the sentence is indented then it is hard to spot.
Our algorithm then runs as follows:
\begin{enumerate}
    \item We compute the distance sequence in $O(n)$ time,
    by incrementally adding $+(\den-\num)$ or $-\num$ as seen in
    \alg{DistMap} and \alg{DistSort}.

    \item We compute the minimal positions
    $a_1,a_2,\ldots,a_p$ and the maximal positions
    $b_1,b_2,\ldots,b_q$ in $O(n)$ time as described above.

    \item \label{en-sweep}
    For each minimal position $a_i$, we find the largest
    maximal position $b_j$ for which $d_{a_i} \leq d_{b_j}$.
    This gives a substring of density $\geq \theta$ and
    length $b_j - a_i + 1$, and we compare this with the longest such
    substring found so far.
\end{enumerate}

The key observation is that,
because minimal and maximal positions are ordered by distance,
step~\ref{en-sweep} can also be performed in $O(n)$ time.
Specifically, if the minimal position $a_i$ is matched with the maximal
position $b_j$, then the next minimal position $a_{i+1}$ will be matched
with \emph{an equal or later maximal position}, i.e., one of
$b_j,b_{j+1},\ldots,b_q$.  We can therefore keep a pointer into the
sequence of maximal positions and slowly move it forward as we process
each of $a_1,\ldots,a_p$, giving step~\ref{en-sweep}
an $O(n)$ running time in total.

\begin{figure}[htb]
    \centering
    \setlength{\fboxsep}{0.5\baselineskip} % The default margin is *very* thin.
    \framebox{\begin{minipage}{0.9\textwidth}\small%
    \begin{algorithmic}[0]
        \Procedure{PositionSweep}{$x_1,\ldots,x_n,\ \theta=\num/\den$}
        \State $d_0 \gets 0$ \Comment{Compute the distance sequence}
        \For{$i \gets 1$ \textbf{to}\ $n$}
            \If{$x_i = 1$}
                \State $d_i \gets d_{i-1} + (\den-\num)$
            \Else
                \State $d_i \gets d_{i-1} - \num$
            \EndIf
        \EndFor

        \Statex

        \State $p \gets 1$\quad;\quad$a_1 \gets 0$
            \Comment{Compute minimal positions}
        \For{$i \gets 1$ \textbf{to}\ $n$}
            \If{$d_i < d_{a_p}$}
                \State $p \gets p + 1$\quad;\quad$a_p \gets i$
            \EndIf
        \EndFor

        \State $q \gets 1$\quad;\quad$b_1 \gets n$
            \Comment{Compute maximal positions}
        \For{$i \gets n-1$ \textbf{downto}\ $0$}
            \If{$d_i > d_{b_q}$}
                \State $q \gets q + 1$\quad;\quad$b_q \gets i$
            \EndIf
        \EndFor

        \Statex

        \State $(a,b) \gets (0,0)$ \Comment{Best start/end found so far}
        \State $j \gets 1$
        \For{$i \gets 1$ \textbf{to}\ $p$}
                \Comment{Run through minimal positions}
            \While{$j < q$ and $d_{a_i} \leq d_{b_{j+1}}$}
                    \Comment{Find best maximal position}
                \State $j \gets j+1$
            \EndWhile
            \If{$b_j - a_i > b - a + 1$}
                \State $(a,b) \gets (a_i + 1,b_j)$
                    \Comment{Longest substring found so far}
            \EndIf
        \EndFor

        \Statex

        \State Output $(a,b)$
        \EndProcedure
    \end{algorithmic}
    \end{minipage}}
    \caption{The \alg{PositionSweep} algorithm for the bounded density problem}
    \label{fig-positionsweep}
\end{figure}

We name this algorithm \alg{PositionSweep}; see
Figure~\ref{fig-positionsweep} for the pseudocode.
Through the discussion above we obtain the following final result:

\begin{lemma}
    The algorithm \alg{PositionSweep} solves the bounded density problem in
    $O(n)$ time in the worst case.
\end{lemma}

\section{Measuring Performance} \label{s-performance}

We finish this paper with a practical field test of the different
algorithms for the fixed density problem.\footnote{We omit the
bounded density problem from this field test because the linear
algorithm \alg{PositionSweep} is simple and slick, with neither the
complexity nor the potential overhead of \alg{DistMatrix}.}
In particular, because
the linear \alg{DistMatrix} algorithm involves a complex data structure with
potentially significant overhead, it is useful to compare its practical
performance against the log-linear but much simpler algorithms
\alg{DistMap} and \alg{DistSort}.
The tests are designed as follows:
\begin{itemize}
    \item We use bitstreams of length $n=10^8$ for all tests.
    This value of $n$ was chosen to be large but manageable.
    We keep $n$ fixed merely to simplify the data presentation---additional
    data has been collected for several smaller values of $n$, and the
    results show similar characteristics to those described here.

    \item All bitstreams are pseudo-random.\footnote{Bitstreams were
    generated using the \texttt{rand()} function from the Linux C~Library.}
    This is of particular benefit to the \alg{SkipMisMatch} algorithm, whose
    expected running time of $O(n \log n)$ in a random scenario is
    significantly better than its worst case time of $O(n^2)$.

    \item We run tests with several different values of the target
    density $\theta$.  This includes values close to and far away from
    $\frac12$, as well as values with small and large denominators---our
    aim is to identify to what degree the performance of different
    algorithms depends upon $\theta$.  The values of $\theta$ that we
    use are $\frac12$, $\frac13$, $\frac25$, $\frac15$,
    $\frac{50}{101}$, $\frac{31}{101}$ and $\frac{1}{101}$.

    \item Each test involves the same 200 pre-generated
    bitstreams of length $n = 10^8$.
    For each algorithm and each value of $\theta$ we measure the mean
    running time over all 200 bitstreams.
    All running times are measured as
    $\mathrm{user} + \mathrm{system}$ time, running on a single 3\,GHz
    Intel Core~2 CPU with 4GB of RAM.
    All algorithms are coded in {\cpp} under GNU/Linux.
\end{itemize}

\begin{figure}[htb]
    \centering
    \includegraphics[scale=0.6]{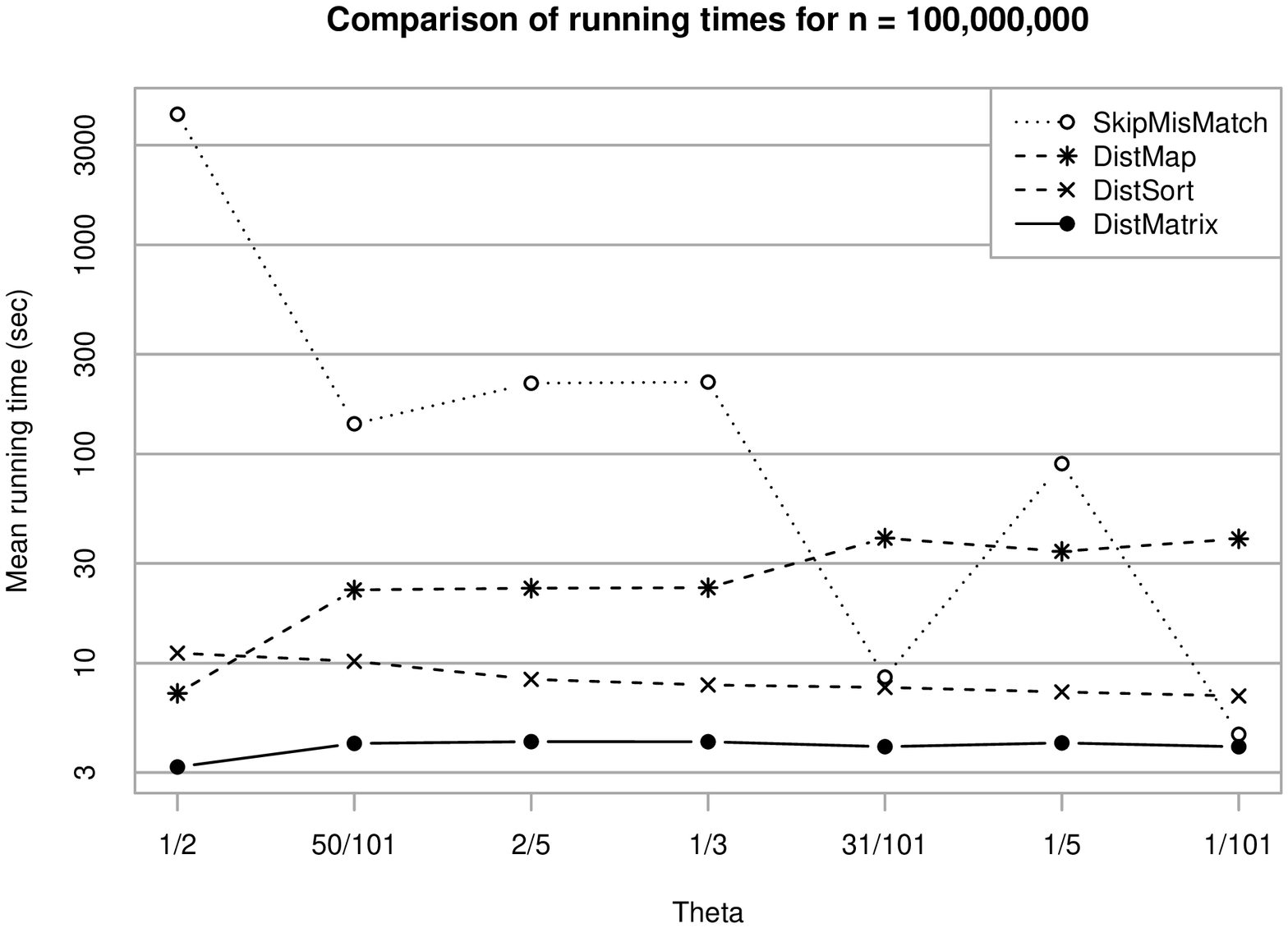}
    \caption{Running times of different algorithms
        for the fixed density problem}
    \label{fig-timesall}
\end{figure}

The results are plotted in Figure~\ref{fig-timesall}; note that
the time axis uses a log scale, with each horizontal line
representing a factor of approximately $\times 3$.
Error bars are not included
because most standard errors are within $\pm 1\%$; the only
exceptions are for $\theta=\frac12$, where \alg{DistMap} has a standard
error of $\pm 1.6\%$ and \alg{SkipMisMatch} has a standard error of
$\pm 10\%$.  The values of $\theta$ are ordered by distance from $\frac12$.

Happily, the results are what we hope for.  The log-linear algorithms
\alg{DistMap} and \alg{DistSort} perform significantly better than
\alg{SkipMisMatch} in most cases, and the linear algorithm \alg{DistMatrix}
consistently outperforms all of the others.

The dependency of \alg{SkipMisMatch} upon $\theta$ is
evident---performance is best when both $|\theta-\frac12|$ and
the denominator $\den$ are large (as expected from
Lemma~\ref{l-skipmismatch-expected}), bringing it close to the 4~second
running time of \alg{DistMatrix} for the extreme case $\theta=\frac{1}{101}$.
At the other extreme, for $\theta=\frac12$ the \alg{SkipMisMatch}
algorithm runs orders of magnitude slower, with a mean running time of
over an hour and some individual cases taking up to $10\frac12$~hours.

\begin{figure}[htb]
    \centering
    \includegraphics[scale=0.6]{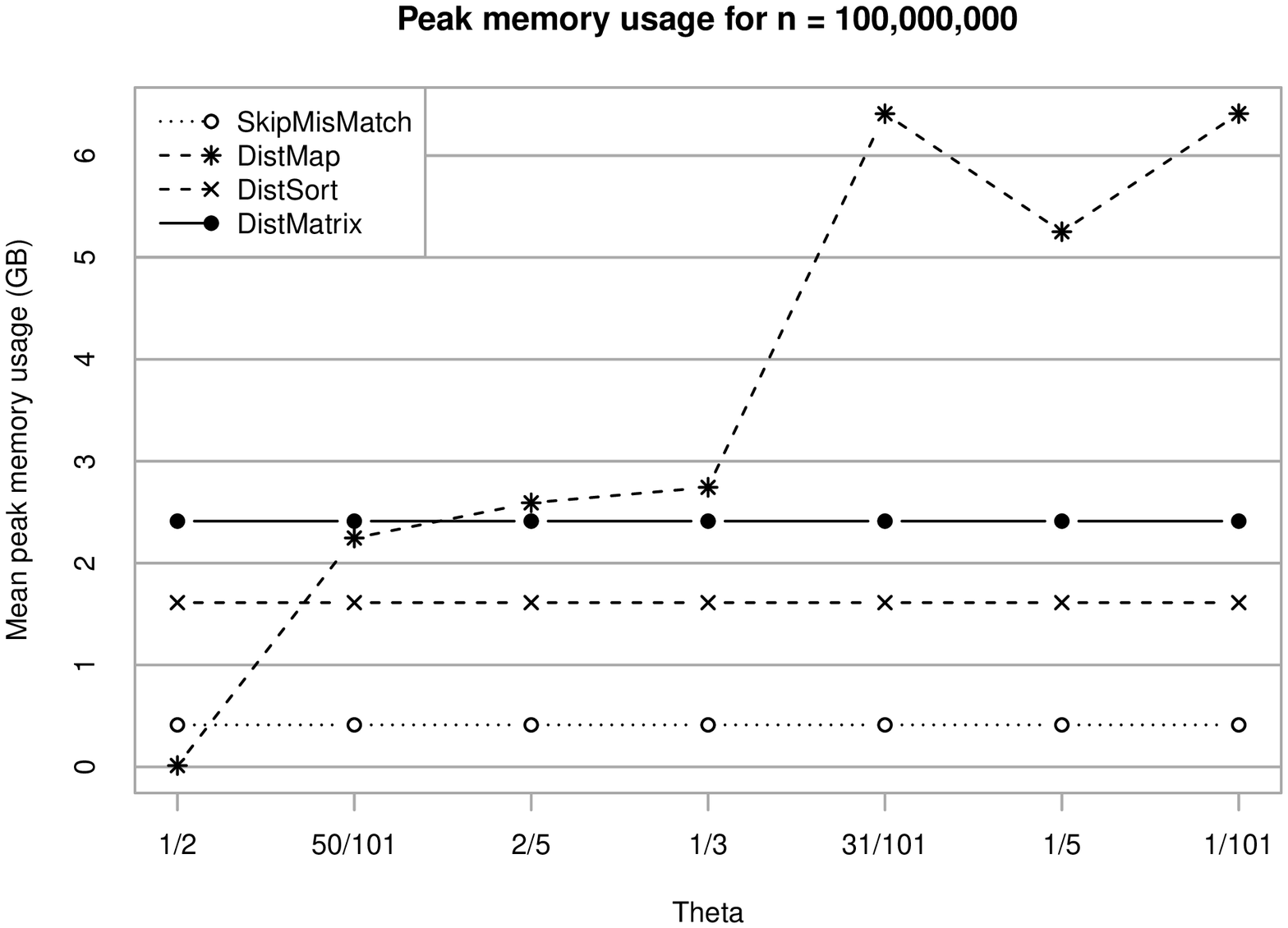}
    \caption{Peak memory usage of different algorithms
        for the fixed density problem}
    \label{fig-memall}
\end{figure}

Amongst the log-linear algorithms\footnote{For \alg{DistMap} and
\alg{DistSort}, the map and sort are implemented using
\texttt{std::map} and \texttt{std::sort} from the {\cpp} Standard Library,
as implemented by the GNU {\cpp} compiler version 4.3.2.},
we find that \alg{DistSort} performs
noticeably better than \alg{DistMap}.  Part of the reason is
the memory overhead due to the map structure---it was found that
\alg{DistMap} often exceeded the available memory on the machine,
burdening it with a reliance on virtual memory (which of course is much
slower).  The linear \alg{DistMatrix} algorithm also suffers from memory
problems to a lesser extent, but Figure~\ref{fig-timesall} shows that
that the effectiveness of the algorithm more than compensates for this.
Figure~\ref{fig-memall} plots the peak memory usage
for each algorithm, again averaged over all 200 bitstreams.

An interesting feature of the running times is that
\alg{DistMap} depends upon $\theta$ in an opposite
manner to \alg{SkipMisMatch}.  This is because when
$\theta \simeq \frac12$ or the denominator $\den$ is small,
there are fewer \emph{distinct} distances amongst $d_0,\ldots,d_n$,
and hence fewer elements stored in the map.

In conclusion, it is pleasing to note how consistently \alg{DistMatrix}
performs across all of the tested values of $\theta$, with mean running
times ranging from $3.2$~seconds to $4.2$~seconds and standard errors of
just $0.1\%$.  The experiments therefore suggest that the
added complexity and overhead of \alg{DistMatrix} are well justified
by the efficiency of the algorithm and its underlying data structure.

%%%%%%%%%%%%%%%%%%%%%%%%%%%%%%%%%%%%%%%%%%%%%%%%%%%%%%%%%%%%%%%%%%%%%%%%
%
%   Acknowledgements
%
%%%%%%%%%%%%%%%%%%%%%%%%%%%%%%%%%%%%%%%%%%%%%%%%%%%%%%%%%%%%%%%%%%%%%%%%

\section*{Acknowledgements}

The author is supported by the Australian Research
Council's Discovery Projects funding scheme (project DP1094516).
He is grateful to Serdar {\boztas}, Mathias Hiron and
Casey Pfluger for fruitful discussions relating to this work.

%%%%%%%%%%%%%%%%%%%%%%%%%%%%%%%%%%%%%%%%%%%%%%%%%%%%%%%%%%%%%%%%%%%%%%%%
%
%   Bibliography
%
%%%%%%%%%%%%%%%%%%%%%%%%%%%%%%%%%%%%%%%%%%%%%%%%%%%%%%%%%%%%%%%%%%%%%%%%

\small
\bibliographystyle{amsplain}
\bibliography{pure,biology}

\newcommand{\noopsort}[1]{}
\providecommand{\bysame}{\leavevmode\hbox to3em{\hrulefill}\thinspace}
\providecommand{\MR}{\relax\ifhmode\unskip\space\fi MR }
\providecommand{\MRhref}[2]{%
  \href{http://www.ams.org/mathscinet-getitem?mr=#1}{#2}
}
\providecommand{\href}[2]{#2}
\begin{thebibliography}{10}

\bibitem{arratia90-erdosrenyi}
R.~Arratia, L.~Gordon, and M.~S. Waterman, \emph{The {E}rd{\H o}s-{R}\'enyi law
  in distribution, for coin tossing and sequence matching}, Ann. Statist.
  \textbf{18} (1990), no.~2, 539--570.

\bibitem{bernardi00-isochores}
Giorgio Bernardi, \emph{Isochores and the evolutionary genomics of
  vertebrates}, Gene \textbf{241} (2000), no.~1, 3--17.

\bibitem{boztas09-density}
Serdar Bozta{\c s}, Simon~J. Puglisi, and Andrew Turpin, \emph{Testing stream
  ciphers by finding the longest substring of a given density}, Information
  Security and Privacy, Lecture Notes in Comput. Sci., vol. 5594, Springer,
  Berlin, 2009, pp.~122--133.

\bibitem{chen05-dragon}
Kevin Chen, Matt Henricksen, William Millan, Joanne Fuller, Leonie Simpson,
  Ed~Dawson, HoonJae Lee, and SangJae Moon, \emph{Dragon: A fast word based
  stream cipher}, Information Security and Cryptology---{ICISC} 2004, Lecture
  Notes in Comput. Sci., vol. 3506, Springer, Berlin, 2005, pp.~33--50.

\bibitem{cormen01-algorithms}
Thomas~H. Cormen, Charles~E. Leiserson, Ronald~L. Rivest, and Clifford Stein,
  \emph{Introduction to algorithms}, 2nd ed., MIT Press, Cambridge, MA, 2001.

\bibitem{duret95-longgenes}
Laurent Duret, Dominique Mouchiroud, and Christian Gautier, \emph{Statistical
  analysis of vertebrate sequences reveals that long genes are scarce in
  {GC}-rich isochores}, J. Mol. Evol. \textbf{40} (1995), no.~3, 308--317.

\bibitem{erdos70-law}
Paul Erd{\H o}s and Alfr{\'e}d R{\'e}nyi, \emph{On a new law of large numbers},
  J. Analyse Math. \textbf{23} (1970), 103--111.

\bibitem{fullerton01-recombination}
Stephanie~M. Fullerton, Antonio~Bernardo Carvalho, and Andrew~G. Clark,
  \emph{Local rates of recombination are positively correlated with {GC}
  content in the human genome}, Mol. Biol. Evol. \textbf{18} (2001), no.~6,
  1139--1142.

\bibitem{goldwasser05-maxdense}
Michael~H. Goldwasser, Ming-Yang Kao, and Hsueh-I Lu, \emph{Linear-time
  algorithms for computing maximum-density sequence segments with
  bioinformatics applications}, J. Comput. System Sci. \textbf{70} (2005),
  no.~2, 128--144.

\bibitem{greenberg03-heavydense}
Ronald~I. Greenberg, \emph{Fast and space-efficient location of heavy or dense
  segments in run-length encoded sequences}, Computing and Combinatorics,
  Lecture Notes in Comput. Sci., vol. 2697, Springer, Berlin, 2003,
  pp.~528--536.

\bibitem{hardison91-alphaglobin}
Ross Hardison, Dan Krane, David Vandenbergh, Jan-Fang Cheng, James Mansberger,
  John Taddie, Scott Schwartz, Xiaoqiu Huang, and Webb Miller, \emph{Sequence
  and comparative analysis of the rabbit $\alpha$-like globin gene cluster
  reveals a rapid mode of evolution in a $\mbox{G}+\mbox{C}$-rich region of
  mammalian genomes}, J. Mol. Biol. \textbf{222} (1991), no.~2, 233--249.

\bibitem{hsieh08-interval}
Yong-Hsiang Hsieh, Chih-Chiang Yu, and Biing-Feng Wang, \emph{Optimal
  algorithms for the interval location problem with range constraints on length
  and average}, IEEE/ACM Trans. Comput. Biol. Bioinformatics \textbf{5} (2008),
  no.~2, 281--290.

\bibitem{knuth97-art2}
Donald~E. Knuth, \emph{The art of computer programming, {V}ol. 2: Seminumerical
  algorithms}, 3rd ed., Addison-Wesley, Reading, MA, 1997.

\bibitem{lin02-length}
Yaw-Ling Lin, Tao Jiang, and Kun-Mao Chao, \emph{Efficient algorithms for
  locating the length-constrained heaviest segments, with applications to
  biomolecular sequence analysis}, Mathematical Foundations of Computer Science
  2002, Lecture Notes in Comput. Sci., vol. 2420, Springer, Berlin, 2002,
  pp.~459--470.

\bibitem{marsaglia85-view}
G.~Marsaglia, \emph{A current view of random number generators}, Computer
  Science and Statistics: The Interface (L.~Billard, ed.), Elsevier Science,
  Amsterdam, 1985, pp.~3--10.

\bibitem{musser97-introsort}
David~R. Musser, \emph{Introspective sorting and selection algorithms}, Softw.
  Pract. Exper. \textbf{27} (1997), no.~8, 983--993.

\bibitem{sharp95-silence}
Paul~M. Sharp, Michalis Averof, Andrew~T. Lloyd, Giorgio Matassi, and John~F.
  Peden, \emph{{DNA} sequence evolution: The sounds of silence}, Phil. Trans.
  R. Soc. Lond. B \textbf{349} (1995), no.~1329, 241--247.

\bibitem{zoubak96-distribution}
Serguei Zoubak, Oliver Clay, and Giorgio Bernardi, \emph{The gene distribution
  of the human genome}, Gene \textbf{174} (1996), no.~1, 95--102.

\end{thebibliography}

%%%%%%%%%%%%%%%%%%%%%%%%%%%%%%%%%%%%%%%%%%%%%%%%%%%%%%%%%%%%%%%%%%%%%%%%
%
%   Author contact and affiliation
%
%%%%%%%%%%%%%%%%%%%%%%%%%%%%%%%%%%%%%%%%%%%%%%%%%%%%%%%%%%%%%%%%%%%%%%%%

\bigskip
\noindent
Benjamin A.~Burton \\
School of Mathematics and Physics, The University of Queensland \\
Brisbane QLD 4072, Australia \\
(bab@debian.org)

\end{document}